\documentclass[aps,amsmath,amssymb,superscriptaddress,prb,preprint]{revtex4-1}

\usepackage[dvipdfmx]{graphicx}

\usepackage{bm}
\usepackage{color} 
\usepackage{comment}

\usepackage[normalem]{ulem}

\begin{document}

\title{Observation of long-range orbital transport \\and giant orbital torque}

\author{Hiroki Hayashi}
\affiliation{Department of Applied Physics and Physico-Informatics, Keio University, Yokohama 223-8522, Japan}

\author{Daegeun Jo}
\affiliation{Department of Physics, Pohang University of Science and Technology, Pohang 37673,Korea}

\author{Dongwook Go}
\affiliation{Peter Gr\"unberg Institut and Institute for Advanced Simulation, Forschungszentrum J\"ulich and JARA, 52425 J\"ulich, Germany \looseness=-1}
\affiliation{Institute of Physics, Johannes Gutenberg University Mainz, 55099 Mainz, Germany}

\author{Tenghua Gao}
\affiliation{Department of Applied Physics and Physico-Informatics, Keio University, Yokohama 223-8522, Japan}
\affiliation{Keio Institute of Pure and Applied Sciences, Keio University, Yokohama 223-8522, Japan}

\author{Satoshi Haku}
\affiliation{Department of Applied Physics and Physico-Informatics, Keio University, Yokohama 223-8522, Japan}

\author{Yuriy Mokrousov}
\affiliation{Peter Gr\"unberg Institut and Institute for Advanced Simulation, Forschungszentrum J\"ulich and JARA, 52425 J\"ulich, Germany \looseness=-1}
\affiliation{Institute of Physics, Johannes Gutenberg University Mainz, 55099 Mainz, Germany}

\author{Hyun-Woo Lee}
\affiliation{Department of Physics, Pohang University of Science and Technology, Pohang 37673,Korea}

\author{Kazuya Ando\footnote{Correspondence and requests for materials should be addressed to ando@appi.keio.ac.jp}}
\affiliation{Department of Applied Physics and Physico-Informatics, Keio University, Yokohama 223-8522, Japan}
\affiliation{Keio Institute of Pure and Applied Sciences, Keio University, Yokohama 223-8522, Japan}
\affiliation{Center for Spintronics Research Network, Keio University, Yokohama 223-8522, Japan}

\maketitle

\bigskip\noindent
\textbf{Abstract\\
Modern spintronics relies on the generation of spin currents through spin-orbit coupling. 
The spin-current generation has been believed to be triggered by current-induced orbital dynamics, which governs the angular momentum transfer from the lattice to the electrons in solids. The fundamental role of the orbital response in the angular momentum dynamics suggests the importance of the orbital counterpart of spin currents: orbital currents. 
However, evidence for its existence has been elusive. Here, we demonstrate the generation of giant orbital currents and uncover fundamental features of the orbital response. We experimentally and theoretically show that orbital currents propagate over longer distances than spin currents by more than an order of magnitude in a ferromagnet and nonmagnets. Furthermore, we find that the orbital current enables electric manipulation of magnetization with efficiencies significantly higher than the spin counterpart. These findings open the door to orbitronics that exploits orbital transport and spin-orbital coupled dynamics in solid-state devices. 
}

\begin{figure}[tb]
\center\includegraphics[scale=1]{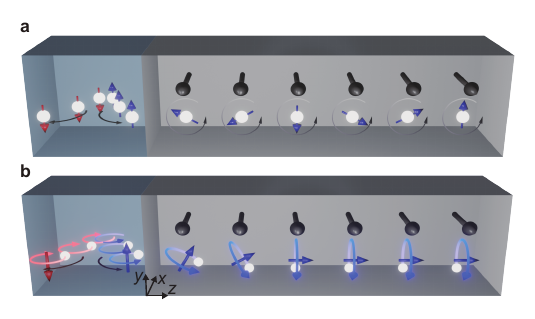}
\caption{
{\bfseries Spin and orbital transport in ferromagnets.} \textbf{a} 
Schematic illustration of the spin Hall effect and spin transport in a ferromagnetic/nonmagnetic bilayer. The black arrow represents the local spins in the ferromagnetic layer. The blue and red arrows denote the up and down spins, respectively. The injected spin, the blue arrow, precesses around the local spin due to the spin exchange coupling in the ferromagnetic layer. \textbf{b} Schematic illustration of the orbital Hall effect and orbital transport in a ferromagnetic/nonmagnetic bilayer. The blue and red arrows, associated with the motion of the electrons, denote the orbital angular momentum. The orbital Hall effect generates the orbital current carrying the $y$ component of the orbital angular momentum $L_y$. The orbital angular momentum, the blue arrow, injected into the ferromagnetic layer induces $z$ component of the orbital angular momentum, $L_z$, through a combined action of the spin-orbit coupling and spin exchange coupling. The induced $L_z$ flows without oscillation through the hot-spots in the momentum space in the ferromagnetic layer. 
}
\label{fig1} 
\end{figure}

\bigskip\noindent
\textbf{Introduction}\\
Since the discovery of the giant magnetoresistance, the concept of spin currents has played a key role in the development of condensed matter physics and spintronics applications~\cite{wolf2001spintronics,Zutic,RevModPhys.80.1517}. Of particular recent interest is the spin Hall effect (SHE), which generates spin currents from charge currents through spin-orbit coupling (see Fig.~\ref{fig1}a)~\cite{hoffmann2013spin,RevModPhys.87.1213}. 
The spin current can interact with local spins, triggering magnetic dynamics in magnetic heterostructures~\cite{RevModPhys.91.035004}. 
The current-induced magnetic dynamics lies at the foundation of a variety of spintronics phenomena, providing a way to realize a plethora of spin-based devices, such as nonvolatile memories, nanoscale microwave sources, and neuromorphic computing devices~\cite{ryu2020current}.

Although spin transport has been central to spintronics, both spin and orbital angular momentum can be carried by electrons in solids~\cite{PhysRevLett.95.066601,PhysRevB.77.165117,PhysRevLett.102.016601,PhysRevLett.121.086602}. An important theoretical prediction is that the SHE is a secondary effect arising from the orbital Hall effect (OHE) in combination with the spin-orbit coupling~\cite{PhysRevLett.102.016601}. The OHE is a phenomenon that generates an orbital current flowing perpendicular to an applied electric field (see Fig.~\ref{fig1}b)~\cite{PhysRevLett.95.066601,PhysRevB.77.165117,PhysRevLett.102.016601,PhysRevLett.121.086602,PhysRevB.98.214405,PhysRevMaterials.5.074407,PhysRevB.101.161409,PhysRevB.101.075429,PhysRevB.102.035409,PhysRevLett.126.056601,PhysRevB.103.085113,PhysRevB.103.195309,PhysRevB.101.121112}, which stems from nonequilibrium interband superpositions of Bloch states with different orbital characters induced by the electric field~\cite{PhysRevLett.121.086602}. This process triggers the transfer of the angular momentum from the lattice to the orbital part of the electron system, and the orbital angular momentum can be further transferred to the spin part of the electrons by the spin-orbit coupling~\cite{PhysRevResearch.2.033401}. This mechanism illustrates the primary role of the orbital response in the angular momentum dynamics in solids, suggesting that the orbital transport is more fundamental than the spin transport. Despite the significance of the orbital response, however, experimental detection of orbital currents remains a major challenge.

The behavior of orbital currents is predicted to be fundamentally different from that of spin currents in ferromagnets (FMs), providing a way to probe the orbital transport~\cite{PhysRevResearch.2.013177,go2021long}. When a spin current is injected into a FM, its transverse component to the magnetization precesses rapidly in the real space because of the spin splitting, which induces the difference in the wavevectors of the majority and minority spins at the Fermi surface (see Fig.~\ref{fig1}a)~\cite{PhysRevB.66.014407}. The precession wavelength is different depending on the incident angles of the electrons, resulting in the short spin decay length, less than 1 nm. In contrast, a recent theory predicts that an orbital current does not precess rapidly and decays over much longer distances than a spin current in FMs does (see Fig.~\ref{fig1}b, the physical picture is explained in Supplementary Note 1)~\cite{go2021long}. This difference has been attributed to the unique feature of the orbital current that its constituent orbital states in FMs can remain nearly degenerate in limited regions of the momentum space, which form hot-spots for the orbital response~\cite{go2021long}. 
In the FMs, the angular momentum of the injected spin and orbital currents is transferred to the local spins, giving rise to torques on the magnetization: spin and orbital torques~\cite{PhysRevResearch.2.013177}.
\textcolor{black}{Although recent experimental studies have suggested the presence of the orbital torque~\cite{PtCo-orbital,PhysRevB.103.L020407,PhysRevResearch.2.013127,tazaki2020current,PhysRevLett.125.177201,Cr-orbital,Ta-orbital,choi2021observation,PhysRevLett.128.067201,PhysRevResearch.4.033037}, the fundamental properties of the orbital torque and orbital transport are still elusive.}
In fact, in the previous works, the observed torque efficiency is much lower than the spin counterpart despite the fundamental role of the orbital response in the angular momentum transport. Furthermore, experimental evidence for the long-range orbital transport in FMs is lacking.

In this work, we report the observation of the long-distance orbital transport and giant orbital torques, revealing the fundamental properties of orbital currents. 
We show that sizable current-induced torques are generated in Ni/Ti bilayers despite the weak spin-orbit coupling of Ti. We find that the torque efficiency increases with increasing the Ni-layer thickness and disappears by replacing the Ni layer with Ni$_{81}$Fe$_{19}$. The unconventional torque, which cannot be attributed to the SHE, is also observed in Ni/W and Ni$_{81}$Fe$_{19}$/W bilayers. We show that these observations are consistent with semirealistic tight-binding calculations, demonstrating that the observed torques originate from the OHE. The experimental and theoretical results evidence that orbital currents propagate over longer distances than spin currents by more than an order of magnitude both in the FM and nonmagnets (NMs). 
Furthermore, we demonstrate that the orbital torque efficiency exceeds the spin torque efficiency of exotic materials, such as topological insulators, as well as Pt, which exhibits the strongest SHE among single element crystals, by an order of magnitude. \textcolor{black}{We also find that the power consumption of the orbital devices can be lower than that of the representative spin-orbitronic devices.} These findings provide unprecedented opportunities for advancing the understanding of the angular momentum dynamics in solids.

\bigskip
\noindent 
\textbf{Results}\\
\textbf{Evidence and characteristics of orbital transport}\\
First, we provide evidence for the existence of orbital currents and orbital torques originating from the OHE in FM/NM structures. To capture the orbital transport, we chose a light metal, Ti, as a source of orbital currents. In Ti, the spin transport plays a minor role because of the weak spin-orbit coupling. In fact, the spin Hall conductivity in Ti is vanishingly small~\cite{du2014systematic}, $\sigma_\mathrm{SH}^\mathrm{Ti}= -1.2$ $(\hbar/e)~\Omega^{-1}$cm$^{-1}$, which is more than three orders of magnitude smaller than the prediction of the intrinsic orbital Hall conductivity $\sigma_\mathrm{OH}^\mathrm{Ti}$: $|\sigma_\mathrm{OH}^\mathrm{Ti}/\sigma_\mathrm{SH}^\mathrm{Ti}|\gg 1$. Another important feature is that the sign of the spin Hall conductivity is opposite to that of the orbital Hall conductivity~\cite{choi2021observation}: $\sigma_\mathrm{OH}^\mathrm{Ti}>0$ and $\sigma_\mathrm{SH}^\mathrm{Ti}<0$. These distinct differences between the SHE and OHE make it possible to distinguish between the spin and orbital Hall currents from the magnitude and sign of the current-induced torques. 
We also note that a recent study reports the detection of current-induced orbital accumulation in Ti by an optical technique~\cite{choi2021observation}. 
These facts make this light metal a promising platform for revealing the fundamental properties of the orbital transport and orbital torques.

To investigate the orbital transport, we measure current-induced torques for Ni/Ti and Ni$_{81}$Fe$_{19}$/Ti films using spin-torque ferromagnetic resonance (ST-FMR). In these heterostructures, as shown in Fig.~\ref{fig1}b, the OHE in the Ti layer generates an orbital Hall current, $\sim k_z L_y$, carrying the $y$ component of the orbital angular momentum, $L_y$, by an electric field applied along the $x$ direction, where $k_z$ is the $z$ component of the wavevector. When the orbital current is injected into the FM layer, the orbital angular momentum interacts with the local spins along the $x$ direction by a combined action of the spin-orbit coupling and spin exchange coupling between the conduction-electron spins and local spins, inducing the $z$ component of the orbital angular momentum, $L_z$. The induced $L_z$ propagates in the FM layer without oscillation through the degenerate orbital hot spots in the momentum space. The propagating $L_z$ interacts with the local spins at each site by a combined action of the spin-orbit coupling and spin exchange coupling, exerting a damping-like (DL) orbital torque on the magnetization in the FM layer (for details, see Supplementary Note 1)~\cite{go2021long}. 
This process indicates that the generation of the orbital torque relies on the spin-orbit coupling in the FM layer, as well as the OHE in the NM layer. Thus, in the presence of both SHE and OHE, the DL-torque efficiency per unit electric field can be expressed as $\xi_\mathrm{DL}^E=T^\mathrm{int}_\mathrm{SH}\sigma_\mathrm{SH}^\mathrm{NM}+\eta_\mathrm{FM}T^\mathrm{int}_\mathrm{OH}\sigma_\mathrm{OH}^\mathrm{NM}$, where $\sigma_\mathrm{SH(OH)}^\mathrm{NM}$ is the spin(orbital) Hall conductivity in the NM layer and $T^\mathrm{int}_\mathrm{SH(OH)}$ is the spin(orbital) transparency at the FM/NM interface. Here, $\eta_\mathrm{FM}$ represents the effective coupling between the orbital angular momentum and magnetization originating from the spin-orbit coupling and spin exchange coupling in the FM layer.

Since $\eta_\mathrm{FM}$ originates from the orbital-to-spin conversion due to the spin-orbit correlation near the Fermi energy in the FM layer~\cite{PhysRevResearch.2.013177}, the orbital torque is sensitive to the electronic structure of the FM layer~\cite{PhysRevResearch.2.033401}.   
Among the conventional 3$d$ FMs, Ni is predicted to show the strongest orbital-to-spin conversion~\cite{Ta-orbital}. 
In the following, we assume that $\eta_\mathrm{FM}$ in Ni$_{81}$Fe$_{19}$ is much weaker than that in Ni: $|\eta_\mathrm{Ni}/\eta_\mathrm{\text{Ni}_{81}\text{Fe}_{19}}|\gg 1$. This assumption is supported by the fact that the physical origin of the strong orbital-to-spin conversion in Ni is in the optimal electronic occupation of $d$ orbital shells such that the Fermi energy is located in the energy gap induced by the spin-orbit coupling~\cite{PhysRevResearch.2.033401}. This is manifested by the strong SHE in Ni, which results from the orbital-to-spin conversion as a result of the combined effect of the OHE and the spin-orbit coupling in the same material~\cite{PhysRevB.98.214405}. The previous work has shown that the spin Hall conductivity in Ni exhibits a sharp spike at the Fermi energy, and the value drops significantly even if the Fermi energy is slightly varied~\cite{PhysRevB.98.214405}. This implies that in a situation like Ni$_{81}$Fe$_{19}$, the efficiency of the orbital-to-spin conversion can be strongly affected by the change of the electronic occupation by Fe doping. Under the assumption of $|\eta_\mathrm{Ni}/\eta_\mathrm{\text{Ni}_{81}\text{Fe}_{19}}|\gg 1$ with the theoretical prediction of $|\sigma_\mathrm{OH}^\mathrm{Ti}/\sigma_\mathrm{SH}^\mathrm{Ti}|\gg 1$, the orbital transport and orbital torque are expected to be pronounced in the Ni/Ti bilayer.

\begin{figure*}[tb]
\center\includegraphics[scale=1]{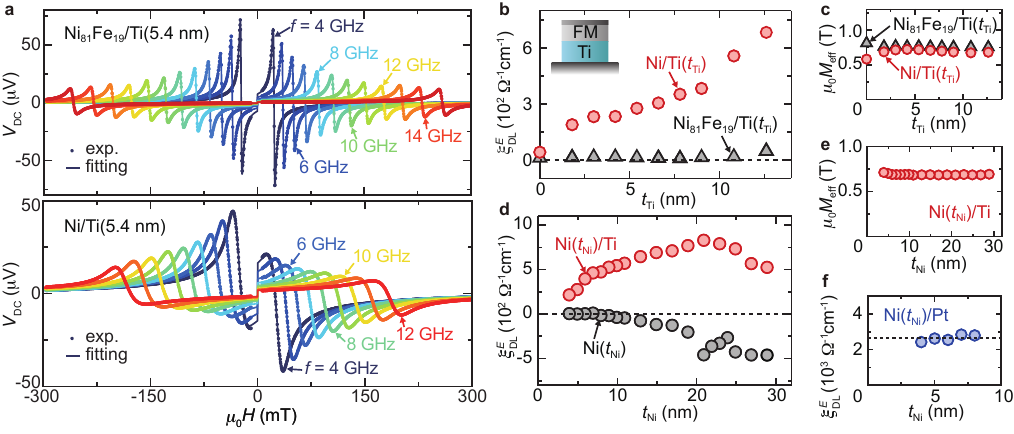}
\caption{
{\bfseries Current-induced torque generated by Ti.}  \textbf{a} Magnetic field $H$ dependence of the DC voltage $V_{\rm DC}$ for the Ni$_{81}$Fe$_{19}$(5~nm)/Ti(5.4~nm) (upper) and Ni(8~nm)/Ti(5.4~nm) (lower) films with an applied RF power of 100~mW at different frequencies $f$. The solid circles are the experimental data and the solid curves are the fitting result. \textbf{b} Ti-layer-thickness $t_\mathrm{Ti}$ dependence of the DL-torque efficiency per unit electric field $\xi_\mathrm{DL}^E$ for the Ni(8~nm)/Ti($t_\mathrm{Ti}$) (red) and Ni$_{81}$Fe$_{19}$(5~nm)/Ti($t_\mathrm{Ti}$) (black) bilayers. Error bars, one-standard-deviation uncertainties from the fitting, are smaller than the symbols. The negligibly small but nonzero positive torque efficiency at $t_\mathrm{Ti} = 0$~nm is consistent with the result for the Ni single-layer film within the experimental uncertainty due to random error. \textbf{c} Ti-layer-thickness $t_\mathrm{Ti}$ dependence of the effective demagnetization field $M_\mathrm{eff}$ for the Ni(8~nm)/Ti($t_\mathrm{Ti}$) (red) and Ni$_{81}$Fe$_{19}$(5~nm)/Ti($t_\mathrm{Ti}$) (black) bilayers. $M_\mathrm{eff}$ was determined from the $f$ dependence of the resonance field $H_\mathrm{res}$ using the Kittel formula: $(2\pi f/\gamma) = \sqrt{\mu_{0}H_\text{res}(\mu_{0}H_\text{res} + \mu_{0}M_\text{eff})}$, where $\gamma$ is the gyromagnetic ratio. \textbf{d} Ni-layer-thickness $t_\mathrm{Ni}$ dependence of $\xi_\mathrm{DL}^E$ for the Ni($t_\mathrm{Ni}$)/Ti(8~nm) bilayer (red) and Ni($t_\mathrm{Ni}$) (black) single-layer films. We have confirmed that the thickness of a magnetic dead layer, $t_\mathrm{dead}$, is less than half a nanometer by measuring $t_\mathrm{Ni}$ dependence of the magnetic moment per unit area for the Ni($t_\mathrm{Ni}$)/Ti films. This result shows that $t_\mathrm{dead}$ is more than an order of magnitude smaller than $t_\mathrm{Ni}$. Thus, the effective thickness of the Ni layer is $t_\mathrm{Ni}-t_\mathrm{dead} \simeq t_\mathrm{Ni}$; the magnetic dead layer can be neglected in the analysis of the DL-torque efficiencies. Here, the range of $t_\mathrm{Ni}$ is above the critical thickness, 1.9 nm, for the amorphous-to-crystalline transition of Ni on Ti~\cite{li2013ni}. \textbf{e} $t_\mathrm{Ni}$ dependence of $M_\mathrm{eff}$ for the Ni($t_\mathrm{Ni}$)/Ti(8~nm) bilayer. \textbf{f} Ni-layer-thickness $t_\mathrm{Ni}$ dependence of $\xi_\mathrm{DL}^E$ for the Ni($t_\mathrm{Ni}$)/Pt(8~nm) bilayer. 
}
\label{fig2} 
\end{figure*}

Figure~\ref{fig2}a shows ST-FMR spectra for the Ni$_{81}$Fe$_{19}$/Ti and Ni/Ti bilayers, measured by applying a radio-frequency (RF) current with a frequency of $f$ and an external magnetic field $H$ (see Methods). The measured spectra are consistent with the prediction of the direct-current (DC) voltage due to the ST-FMR:~\cite{PhysRevLett.106.036601,fang2011spin}
\begin{equation}
V_{\rm DC}=V_{\rm sym}\frac{W^2}{(\mu_0 H-\mu_0H_{\rm res})^2+W^2}+V_{\rm antisym}\frac{W(\mu_0 H-\mu_0H_{\rm res})}{(\mu_0 H-\mu_0H_{\rm res})^2+W^2},
\label{eq:VDC}
\end{equation}
where $W$ is the linewidth and $ H_{\rm res}$ is the FMR field. 
Here, the symmetric component $V_{\rm sym}$ is proportional to the DL effective field $H_{\rm DL}$, while the antisymmetric component $V_{\rm antisym}$ is proportional to the sum of the Oersted field $H_{\rm Oe}$ and field-like effective field $H_{\rm FL}$. 
We confirmed that magnetic field angle dependence of $V_{\rm sym}$ and $V_{\rm antisym}$ is consistent with the prediction of the ST-FMR model (see Supplementary Note 2). Notable is that the ST-FMR spectral shape is clearly different between the Ni$_{81}$Fe$_{19}$/Ti and Ni/Ti bilayers. 
In the ST-FMR spectra for the Ni$_{81}$Fe$_{19}$/Ti bilayer, the symmetric component $V_{\rm sym}$ is vanishingly small, consistent with previous reports that demonstrate negligible DL torque in Ti-based structures~\cite{PhysRevB.93.180402,PhysRevApplied.15.L031001}. 
In contrast, $V_{\rm sym}$ is clearly observed for the Ni/Ti bilayer, demonstrating the generation of a sizable DL torque in this system. Here, the observed $V_\mathrm{sym}$ signals cannot be attributed to spin-pumping and thermoelectric signals (see Supplementary Note 2).

In Fig.~\ref{fig2}b, we show Ti-layer thicknesses $t_\mathrm{Ti}$ dependence of the DL-torque efficiency per unit electric field, $\xi_\text{DL}^E = \zeta (2e/\hbar )\mu_{0}M_\text{s}t_\text{FM}H_\text{DL}/E$, determined from the ST-FMR for the Ni$_{81}$Fe$_{19}$/Ti and Ni/Ti bilayers, where $M_\mathrm{s}$ is the saturation magnetization, $t_\text{FM}$ is the thickness of the FM layer, and $E$ is the applied electric field (for details, see Supplementary Note 3). Here, $\zeta=-1$ for FM/NM/substrate structures and $\zeta=1$ for NM/FM/substrate structures. 
Figure~\ref{fig2}b shows that $\xi_\text{DL}^E$ of the Ni/Ti bilayer increases with increasing $t_\mathrm{Ti}$. We confirmed that the variation in $\xi_\text{DL}^E$ with $t_\mathrm{Ti}$ is not induced by a possible change of the magnetic property of the Ni layer; the effective demagnetization field $M_\mathrm{eff}$ is independent of $t_\mathrm{Ti}$, as shown in Fig.~\ref{fig2}c.
Thus, the clear increase in $\xi_\text{DL}^E$ with $t_\mathrm{Ti}$ indicates that the observed DL-torque originates from the bulk effects, the SHE or OHE, in the Ti layer; self-induced torques in the Ni layer and interfacial spin-orbit coupling effects are not the source of the observed DL torque (see also Supplementary Notes 3 and 4). The negligible contribution from the interfacial spin-orbit coupling is consistent with a recent study~\cite{PhysRevApplied.15.L031001}. We also note that the sizable DL torque in the Ni/Ti bilayer is supported by second-harmonic Hall resistance measurements (see Supplementary Note 5).

The unconventional torque in the Ni/Ti bilayer is consistent with the orbital torque originating from the OHE in the Ti layer. We note that the sign of the DL-torque in the Ni/Ti layer is opposite to the prediction of the SHE but is consistent with that of the OHE in the Ti layer. 
Furthermore, the observed DL-torque efficiency is more than two orders of magnitude higher than the spin Hall conductivity of Ti. These results provide clear evidence that the SHE in the Ti layer is not responsible for the observed DL torque. 
Figure~\ref{fig2}b also shows that the DL-torque efficiency $\xi_\text{DL}^E $ of the Ni/Ti bilayer is more than an order of magnitude larger than that of the Ni$_{81}$Fe$_{19}$/Ti bilayer, demonstrating that the electronic structure of the FM layer plays a crucial role in generating the observed torque. 
The distinct difference in the DL-torque efficiency between the Ni/Ti and Ni$_{81}$Fe$_{19}$/Ti devices is consistent with the scenario of the orbital torque with the assumption of $|\eta_\mathrm{Ni}/\eta_\mathrm{\text{Ni}_{81}\text{Fe}_{19}}|\gg 1$. 
\textcolor{black}{We also note that the orbital transparency $T_\mathrm{OH}^\mathrm{int}$ can also be different between the Ni/Ti and Ni$_{81}$Fe$_{19}$/Ti devices.}
Since the spin Hall conductivity in Ti is vanishingly small~\cite{du2014systematic}, $\sigma_\mathrm{SH}^\mathrm{Ti}= -1.2$ $(\hbar/e)~\Omega^{-1}$cm$^{-1}$, the DL torque due to the SHE is negligible regardless of $t_\mathrm{Ti}$, resulting in the OHE-dominated torque with $\xi_\mathrm{DL}^E>0$ over the thickness range investigated in the Ni/Ti bilayer.
We also demonstrate magnetization switching by the orbital torque (see Supplementary Note 6).

We find that the DL-torque efficiency in the Ni/Ti bilayer is further enhanced by increasing the Ni-layer thickness $t_\mathrm{Ni}$, demonstrating the long-range orbital transport in the Ni layer. 
Figure~\ref{fig2}d shows that $\xi_\mathrm{DL}^E$ increases with increasing $t_\mathrm{Ni}$ up to $t_\mathrm{Ni}=20$~nm despite the fact that the magnetic property is unchanged as shown in Fig.~\ref{fig2}e. 
In the scenario of spin torques, since spin currents decay within 1~nm due to the spin dephasing, the DL-torque efficiency is independent of the FM-layer thickness $t_\mathrm{FM}$ when $t_\mathrm{FM}>1$~nm. In fact, we confirmed that $\xi_\mathrm{DL}^E$ is independent of $t_\mathrm{Ni}$ in Ni/Pt bilayers, where the DL-torque is dominated by the SHE in the Pt layer, as shown in Fig.~\ref{fig2}f (see also Supplementary Note 7). In contrast, since orbital currents can propagate over much longer distances than the spin dephasing length in FMs, the DL-torque efficiency increases with $t_\mathrm{FM}$~\cite{go2021long}. The observed $t_\mathrm{Ni}$ dependence of $\xi_\mathrm{DL}^E$ demonstrates that the orbital currents responsible for the DL torque propagate over longer distances than the spin dephasing length by an order of magnitude in the Ni layer. Here, the suppression of $\xi_\mathrm{DL}^E$ in the Ni/Ti device when $t_\mathrm{Ni}>20$~nm can be attributed to a self-induced torque in the Ni layer. As shown in Fig.~\ref{fig2}d, the DL-torque, whose sign is opposite to that of the Ni/Ti bilayer, is non-negligible in a Ni single-layer film when $t_\mathrm{Ni}>10$~nm. This result is consistent with the scenario of the self-induced torque, which is non-negligible only when the magnetic layer is thicker than the exchange length (8.4~nm for Ni)~\cite{WangASOT}.
In contrast to the self-induced torque, which increases with $t_\mathrm{Ni}$ and becomes sizable especially at $t_\mathrm{Ni} > 20$~nm, the orbital torque tends to saturate with increasing $t_\mathrm{Ni}$. The different $t_\mathrm{Ni}$ dependences result in the suppression of $\xi_\mathrm{DL}^E$ in the Ni/Ti bilayer at $t_\mathrm{Ni}> 20$~nm.

\bigskip
\noindent 
\textbf{Crossover between spin and orbital torques}\\
Next, we investigate the competition between spin and orbital torques by replacing the light metal Ti with a heavy metal W. The two metals are different in terms of the strength of the spin-orbit coupling. 
In the Ti-based device, because of the weak spin-orbit coupling, the spin transport and spin torques are negligible. 
In contrast, in the W-based system, the SHE contributes to the DL torque due to the strong spin-orbit coupling.  
Although the spin and orbital torques coexist in the W-based system, it is still possible to clarify the dominant mechanism of the angular momentum transport, spin or orbital channels, because the sign of the orbital Hall conductivity in W~\cite{PhysRevLett.102.016601}, $\sigma_\mathrm{OH}^{\mathrm{W}}>0$, is opposite to that of the spin Hall conductivity~\cite{PhysRevB.96.241105}, $\sigma_\mathrm{SH}^{\alpha-\mathrm{W}}=-785$~$(\hbar/e)~\Omega^{-1}$cm$^{-1}$ in $\alpha$-W and $\sigma_\mathrm{SH}^{\beta-\mathrm{W}}=-1255$~$(\hbar/e)~\Omega^{-1}$cm$^{-1}$ in $\beta$-W.

\begin{figure}[tb]
\center\includegraphics[scale=1]{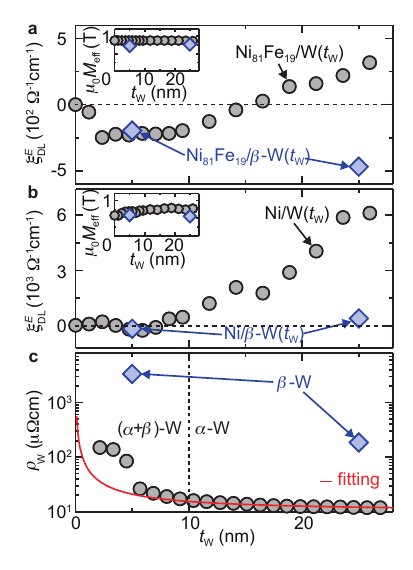}
\caption{
{\bfseries Current-induced torque generated by W.}  \textbf{a} W-layer-thickness $t_\mathrm{W}$ dependence of $\xi_\mathrm{DL}^E$ for the Ni$_{81}$Fe$_{19}$(5~nm)/W($t_\mathrm{W}$) (black circles) and Ni$_{81}$Fe$_{19}$(5~nm)/$\beta$-W($t_\mathrm{W}$) (blue diamonds) bilayers. The inset shows $t_\mathrm{W}$ dependence of $M_\mathrm{eff}$. Error bars, one-standard-deviation uncertainties from the fitting, are smaller than the symbols. Here, $\xi_\mathrm{DL}^E$ (blue diamonds) of the Ni$_{81}$Fe$_{19}$/$\beta$-W with $t_\mathrm{W} = 5$~nm is smaller than that of the Ni$_{81}$Fe$_{19}$/$\beta$-W with $t_\mathrm{W} = 25$~nm, even though both torques are dominated by the spin Hall effect in the $\beta$-W layer. This difference can be attributed to the suppression of the intrinsic spin Hall conductivity in the dirty-metal regime due to the shortening of the carrier lifetime induced by decreasing $t_\mathrm{W}$. This interpretation is supported by the fact that the resistivity of the $\beta$-W film increases with decreasing the thickness from $t_\mathrm{W} = 25$~nm to $t_\mathrm{W} = 5$~nm, as shown in Fig.~\ref{fig3}c. 
\textbf{b} $t_\mathrm{W}$ dependence of $\xi_\mathrm{DL}^E$ for the Ni(8~nm)/W($t_\mathrm{W}$) (black circles) and Ni(8~nm)/$\beta$-W($t_\mathrm{W}$) (blue diamonds) bilayers. \textbf{c} $t_\mathrm{W}$ dependence of the resistivity $\rho_\mathrm{W}$ for the W (black circles) and $\beta$-W (blue diamonds) films. The resistivity shows that the W film (black circles) is $\alpha$-W when $t_\mathrm{W}>10$~nm and the mixture of $\alpha$- and $\beta$-phase W when $t_\mathrm{W}<10$~nm. The red curve is the fitting result using $\rho_\mathrm{W}(t_\mathrm{W})=a t_\mathrm{W}^{-1}+\rho_\mathrm{W}^{\text{bulk}}$, where $a t_\mathrm{W}^{-1}$ represents the resistivity due to the surface scattering and $\rho_\mathrm{W}^{\text{bulk}}$ is the resistivity in the bulk limit. 
}
\label{fig3} 
\end{figure}

In Figs.~\ref{fig3}a and \ref{fig3}b, we show W-thickness $t_\mathrm{W}$ dependence of the DL-torque efficiency $\xi_\mathrm{DL}^E$, determined by the ST-FMR, for Ni$_{81}$Fe$_{19}$/W and Ni/W films (see the black circles and also Supplementary Notes 2 and 3). We first focus on $\xi_\mathrm{DL}^E$ around $t_\mathrm{W}=20$~nm, where $\xi_\mathrm{DL}^E>0$ in both Ni$_{81}$Fe$_{19}$/W and Ni/W films. Figures~\ref{fig3}a and \ref{fig3}b show that $\xi_\mathrm{DL}^E$ of the Ni/W bilayer is more than an order of magnitude higher than that of the Ni$_{81}$Fe$_{19}$/W bilayer, and $\xi_\mathrm{DL}^E$ increases with $t_\mathrm{W}$ in both Ni$_{81}$Fe$_{19}$/W and Ni/W bilayers. These unconventional features are consistent with the scenario of the orbital torque, observed for the FM/Ti devices (see also Supplementary Note 8).
In fact, the sign of the DL torque, $\xi_\mathrm{DL}^E>0$, is consistent with the OHE and opposite to the SHE in W, indicating that the DL torque is dominated by the orbital transport around $t_\mathrm{W} = 20$~nm. Here, the validity of the determined DL-torque efficiency is supported by second-harmonic Hall resistance measurements (see Supplementary Note 5).

The observed $t_\mathrm{W}$ dependence of $\xi_\mathrm{DL}^E$ for the FM/W bilayers demonstrates the crossover from the SHE-dominant regime to the OHE-dominant regime, illustrating different length scales of the spin and orbital transport. 
In the Ni$_{81}$Fe$_{19}$/W bilayer, the sign of $\xi_\mathrm{DL}^E$ changes from negative to positive by increasing the W thickness $t_\mathrm{W}$ (see the black circles in Fig.~\ref{fig3}a). The sign of $\xi_\mathrm{DL}^E$ in the Ni/W bilayer, where the OHE contribution is more pronounced due to the larger $\eta_\mathrm{FM}$, also changes from negative to positive around $t_\mathrm{W} = 8$~nm.
In the W-based devices, the spin Hall contribution to the DL torque with $\xi_\mathrm{DL,SH}^E=T^\mathrm{int}_\mathrm{SH}\sigma_\mathrm{SH}^\mathrm{W}<0$ should be saturated when $t_\mathrm{W}$ is larger than the spin diffusion length $\approx 1.5$~nm in W~\cite{PhysRevMaterials.2.014403}. In contrast, the orbital Hall contribution with $\xi_\mathrm{DL,OH}^E=\eta_\mathrm{FM}T^\mathrm{int}_\mathrm{OH}\sigma_\mathrm{OH}^\mathrm{W}>0$ increases as $t_\mathrm{W}$ increases and becomes comparable to the long orbital decay length, as evidenced by the clear increase in $\xi_\mathrm{DL}^E$ with $t_\mathrm{W}$ of around 20~nm in the FM/W bilayers. Because of the different $t_\mathrm{W}$ dependence, the ratio between the orbital and spin torque efficiencies, $|\xi_\mathrm{DL,OH}^E/\xi_\mathrm{DL,SH}^E|=|\eta_\mathrm{FM}T^\mathrm{int}_\mathrm{OH}\sigma_\mathrm{OH}^\mathrm{W}/T^\mathrm{int}_\mathrm{SH}\sigma_\mathrm{SH}^\mathrm{W}|$, increases with $t_\mathrm{W}$. The observed variation in $\xi_\mathrm{DL}^E$ with $t_\mathrm{W}$ in the FM/W bilayers is consistent with the competition between the spin Hall and orbital Hall contributions. In fact, Fig.~\ref{fig3}a shows that $\xi_\mathrm{DL}^E$ is almost saturated around $t_\mathrm{W}=2$~nm, which is comparable to the spin diffusion length in W, and remains so until the orbital Hall contribution shows up at large $t_\mathrm{W}$. We also note that the sign and magnitude of the saturation value of $\xi_\mathrm{DL}^E$ is consistent with the spin Hall conductivity of W,  supporting the dominant role of the SHE in generating the DL torque in the Ni$_{81}$Fe$_{19}$/W device with $t_\mathrm{W}\sim5$~nm.

Figure~\ref{fig3}b demonstrates that the orbital torque efficiency in the Ni/W bilayer is comparable or larger than the DL-torque efficiency originating from the SHE in the Ni/Pt bilayer (see Fig.~\ref{fig2}f). The efficient torque generation is a unique feature of $\alpha$-W. Here, the resistivity $\rho_\mathrm{W}$ of the W layer, shown in Fig.~\ref{fig3}c, indicates that the W layer is low-resistivity $\alpha$-W when $t_\mathrm{W}>10$~nm, while the W layer is a mixture of low-resistivity $\alpha$-phase and high-resistivity $\beta$-phase when $t_\mathrm{W}<10$~nm (for details, see Methods).
To test the role of the structural phase in generating the DL torque, we also fabricated Ni/$\beta$-W and Ni$_{81}$Fe$_{19}$/$\beta$-W films with  $t_\mathrm{W}=5$ and 25 nm by changing the sputtering condition (see the blue diamonds in Figs.~\ref{fig3}a-c and Methods).
Figure~\ref{fig3}b shows that the high DL-torque efficiency of the Ni/$\alpha$-W film at $t_\mathrm{W}=25$~nm is significantly suppressed by replacing the $\alpha$-W layer with the $\beta$-W layer, suggesting that that $|T^\mathrm{int}_\mathrm{OH}\sigma_\mathrm{OH}^\mathrm{W}/T^\mathrm{int}_\mathrm{SH}\sigma_\mathrm{SH}^\mathrm{W}|$ of the $\beta$-W device is much lower than that of the $\alpha$-W device.

The result for the $\beta$-W devices shows $\xi_\mathrm{DL}^E>0$ only in the Ni/$\beta$-W device with $t_\mathrm{W}=25$~nm, which highlights the larger $\eta_\mathrm{FM}$ in Ni, as well as the different $t_W$ dependence of $\xi_\mathrm{DL,SH}^E$ and $\xi_\mathrm{DL,OH}^E$.
In the Ni$_{81}$Fe$_{19}$/$\beta$-W devices, the SHE provides the dominant contribution to $\xi_\mathrm{DL}^E$ because of the following two reasons. First, the ratio $|T^\mathrm{int}_\mathrm{OH}\sigma_\mathrm{OH}^\mathrm{W}/T^\mathrm{int}_\mathrm{SH}\sigma_\mathrm{SH}^\mathrm{W}|$ of the $\beta$-W devices is much lower than that of the $\alpha$-W devices. Second, $\eta_\mathrm{FM}$ of Ni$_{81}$Fe$_{19}$ is smaller than that of Ni. The combination of these two features results in negligible $|\xi_\mathrm{DL,OH}^E/\xi_\mathrm{DL,SH}^E|=|\eta_\mathrm{FM}T^\mathrm{int}_\mathrm{OH}\sigma_\mathrm{OH}^\mathrm{W}/T^\mathrm{int}_\mathrm{SH}\sigma_\mathrm{SH}^\mathrm{W}|$ in the Ni$_{81}$Fe$_{19}$/$\beta$-W devices; the SHE provides the dominant contribution to $\xi_\mathrm{DL}^E$ even at $t_\mathrm{W} = 25$~nm. 
Here, the effective spin Hall angles obtained from $\xi_\mathrm{DL}^E$ for the Ni$_{81}$Fe$_{19}$/$(\alpha+\beta)$-W (black circles) and Ni$_{81}$Fe$_{19}$/$\beta$-W (blue diamonds) devices with $t_\mathrm{W} = 5$~nm are $\theta_\mathrm{SH,eff}^{(\alpha+\beta)\mathrm{-W}}=\xi_\mathrm{DL}^E \rho_{(\alpha+\beta)\mathrm{-W}}=-0.02$ for $(\alpha+\beta)$-W and $\theta_\mathrm{SH,eff}^{\beta\mathrm{-W}}=\xi_\mathrm{DL}^E \rho_{\beta\mathrm{-W}}=-0.65$ for $\beta$-W. These values are consistent with literature~\cite{PhysRevB.96.241105,RevModPhys.91.035004}, supporting the dominant role of the SHE in the Ni$_{81}$Fe$_{19}$/W devices. 
In contrast, the OHE contribution is non-negligible in the Ni/$\beta$-W device due to the larger $\eta_\mathrm{FM}$ of Ni; although the SHE provides the dominant contribution to $\xi_\mathrm{DL}^E$ at $t_\mathrm{W} = 5$~nm, the contribution from the OHE exceeds that from the SHE in the Ni/$\beta$-W device at $t_\mathrm{W} = 25$~nm because $|\xi_\mathrm{DL,OH}^E/\xi_\mathrm{DL,SH}^E|$ increases with $t_\mathrm{W}$ due to the long orbital decay length, resulting in the sign reversal of $\xi_\mathrm{DL}^E$.

\bigskip
\noindent 
\textbf{Semirealistic calculation}\\

\begin{figure}[tb]
	\center\includegraphics[scale=1]{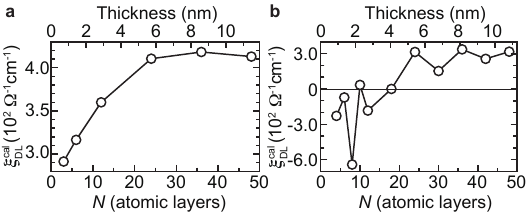}
	\caption{
		{\bfseries Theoretical calculations of the current-induced torque.} Torque efficiency from the theoretical calculations $\xi_\mathrm{DL}^\mathrm{cal}$ for  (\textbf{a}) Ni(12)/Ti($N$) and (\textbf{b}) Ni(12)/W($N$) bilayer structures. $N$ is the number of atomic layers of the nonmagnet and the corresponding thickness is indicated on the top axis.
	}
	\label{fig4} 
\end{figure}

We perform the numerical calculation of the DL torque. The first-principles calculation is limited to very thin systems with the thickness of a few nm. To examine experimental situations with much thicker systems, we combine the first-principles calculation scheme~\cite{fleur} with the tight-binding scheme. The resulting semirealistic calculation scheme~\cite{go2021long} (for details, see Supplementary Notes 9 and 10) allows the torque calculation up to $\sim$ 10 nm thick systems for Ni(12)/Ti($N$) and Ni(12)/W($N$) bilayers (the integers in the parentheses denote the number of atomic layers). For various thicknesses of the nonmagnet, $N$, we obtain the electronic structures and calculate the DL torque under an external electric field $\mathcal{E}_x \hat{\mathbf{x}}$ by evaluating the Kubo formula~\cite{PhysRevResearch.2.033401},
		\begin{equation}\label{eqn:kubo}
		T_\mathrm{DL} = \frac{e\hbar \mathcal{E}_x}{N_\mathbf{k}} \sum_\mathbf{k} \sum_{m \neq n} 
		(f_{n\mathbf{k}} - f_{m\mathbf{k}}) 
		\mathrm{Im}\left[\frac{\langle u_{n\mathbf{k}} \vert (\mathbf{T}_\mathrm{XC})_y \vert u_{m\mathbf{k}} \rangle 
			\langle u_{m\mathbf{k}} \vert v_x(\mathbf{k}) \vert u_{n\mathbf{k}} \rangle}
		{(E_{n\mathbf{k}} - E_{m\mathbf{k}} + i\Gamma )^2}\right],
	\end{equation}
where $e>0$ is the elementary charge, $N_\mathbf{k}$ is the number of $\mathbf{k}$ points used to sample the Brillouin zone, $\vert u_{n\mathbf{k}} \rangle$ is the periodic part of the Bloch state with energy eigenvalue $E_{n\mathbf{k}}$, $f_{n\mathbf{k}}$ is the Fermi-Dirac distribution function, $v_x(\mathbf{k})$ is the $x$-component of the velocity operator, $\Gamma$ is the energy broadening, and $\mathbf{T}_\mathrm{XC} = \frac{2\mu_\mathrm{B}}{\hbar} \mathbf{S} \times \bm{\Omega}_\mathrm{XC} $ is the exchange torque operator with the Bohr magneton $\mu_\mathrm{B}$, spin operator $\mathbf{S}$, and exchange field operator $\mathbf{\Omega_\mathrm{XC}}$. Figure~\ref{fig4}a shows the torque efficiency $\xi_\mathrm{DL}^\mathrm{cal} = (e/\hbar)T_\mathrm{DL}/(A_\mathrm{cell}\mathcal{E}_x)$ ($A_\mathrm{cell}$ is the unit cell area) as a function of $N$ for Ni(12)/Ti($N$). The torque efficiency increases as Ti layer becomes thicker and it saturates at the thickness of approximately 10 nm, where $\xi_\mathrm{DL}^\mathrm{cal} > 400$ $\Omega^{-1}$cm$^{-1}$ is of the same order of magnitude as our experimental values for Ni/Ti (Fig.~\ref{fig2}b). We note that this $\xi_\mathrm{DL}^\mathrm{cal}$ is dominated by the orbital torque while the SHE gives an insignificant contribution (Supplementary Note 9). Figure~\ref{fig4}b shows $\xi_\mathrm{DL}^\mathrm{cal}$ as a function of $N$ for Ni(12)/W($N$) and it exhibits a significant thickness dependence with a characteristic length $\gtrsim 10$ nm which is an order of magnitude longer than the spin diffusion length $\approx 1.5$~nm in W~\cite{PhysRevMaterials.2.014403}. We also find that this long-range behavior stems from the orbital torque which has a positive sign, whereas the spin Hall contribution has a negative sign with a much shorter length scale (Supplementary Note 9). As a result of the competition between OHE and SHE, the sign of the torque changes from negative to positive as the W layer becomes thicker (Fig.~\ref{fig4}b), which is also observed in our experiment (Fig.~\ref{fig3}). Hence, our theoretical calculations provide further evidence of the positive long-range orbital torque and support our experimental observations.

\begin{figure}[tb]
\center\includegraphics[scale=1]{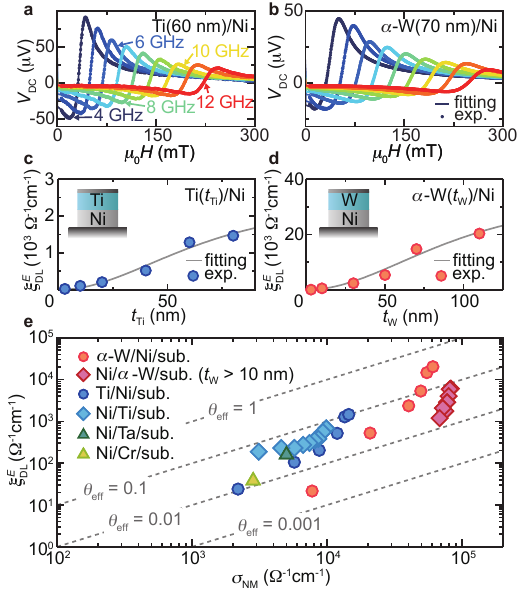}
\caption{
{\bfseries Relation between orbital torque efficiency and electric conductivity.} 
Magnetic field $H$ dependence of the DC voltage $V_{\rm DC}$ for the (\textbf{a}) Ti(60~nm)/Ni(8~nm) and (\textbf{b}) $\alpha$-W(70~nm)/Ni(8~nm) films with an applied RF power of 100~mW at different frequencies $f$. 
\textbf{c} Ti-layer-thickness $t_\mathrm{Ti}$ dependence of $\xi_\mathrm{DL}^E$ for the Ti($t_\mathrm{Ti}$)/Ni(8~nm) bilayer. \textbf{d} W-layer-thickness $t_\mathrm{W}$ dependence of $\xi_\mathrm{DL}^E$ for the $\alpha$-W($t_\mathrm{W}$)/Ni(8~nm) bilayer. The solid circles are the experimental data and the solid curves are the fitting result assuming that the orbital Hall conductivity is independent of $t_\mathrm{NM}$. \textbf{e} The DL-torque efficiency $\xi_\mathrm{DL}^E$ as a function of the longitudinal electric conductivity $\sigma_\mathrm{NM}$ for all the devices investigated in this study. The data for the Ni/Ta and Ni/Cr structures are taken from published papers~\cite{Ta-orbital,Cr-orbital}. The dotted lines are the effective Hall angles $\theta_\mathrm{eff}=\xi_\mathrm{DL}^E/\sigma_\mathrm{NM}$. Error bars, one-standard-deviation uncertainties from the fitting, are smaller than the symbols. }
\label{fig5} 
\end{figure}

\bigskip
\noindent 
\textbf{Orbital torque in clean limit}\\
Finally, we demonstrate exceptionally high orbital torque efficiencies beyond the prediction of the intrinsic OHE. The evidence for this is obtained by further increasing the NM-layer thickness $t_\mathrm{NM}$, a situation that is difficult to capture by semirealistic tight-binding calculations. Figures~\ref{fig5}a and \ref{fig5}b show ST-FMR spectra for Ti(60~nm)/Ni and $\alpha$-W(70~nm)/Ni bilayers, where the stacking order is changed from FM/NM/SiO$_2$-substrate to NM/FM/SiO$_2$-substrate to minimize the interfacial roughness in the thicker devices. Figures~\ref{fig5}a and \ref{fig5}b show that the sign of the ST-FMR voltage for the NM/Ni devices is opposite to that for the Ni/NM devices (see Fig.~\ref{fig2}a), as expected for the reversed stacking order. From the ST-FMR result, we obtain $t_\mathrm{Ti}$ and $t_\mathrm{W}$ dependence of $\xi_\mathrm{DL}^E$, as shown in Figs.~\ref{fig5}c and \ref{fig5}d. The results for $t_\mathrm{NM}<30$~nm are qualitatively consistent with $\xi_\mathrm{DL}^E$ of the Ni/NM bilayers, shown in Figs.~\ref{fig2}b and \ref{fig3}b, supporting that the observed torques are dominated by the OHE. 
Here, the thickness of the Ni layer, 8~nm, in the NM/Ni devices is thin enough to neglect the contribution from the self-induced torque (see also Fig.~\ref{fig2}d).

Figures~\ref{fig5}c and \ref{fig5}d show that $\xi_\mathrm{DL}^E$ increases with $t_\mathrm{NM}$ even in the very thick devices ($t_\mathrm{NM}>30$~nm), revealing the exceptionally high orbital torque efficiencies. When we assume that the orbital Hall conductivity is independent of $t_\mathrm{NM}$, 
the $t_\mathrm{NM}$ dependence of the orbital torque efficiency $\xi_\mathrm{DL}^E=\eta_\mathrm{FM}T^\mathrm{int}_\mathrm{OH}\sigma_\mathrm{OH}^\mathrm{NM}$ can be expressed as $\xi_{\mathrm{DL}}^{E}(t_{\mathrm{NM}} )=\xi_{\mathrm{DL}, 0}^{E}\left[1-\operatorname{sech}\left(t_{\mathrm{NM}} / \lambda_{\mathrm{NM}}\right)\right]$, where $\lambda_\mathrm{NM}$ is the orbital decay length. By fitting the results in Figs.~\ref{fig5}c and \ref{fig5}d, 
we obtain $\lambda_{\mathrm{Ti}}=47\pm11$~nm and $\xi_\mathrm{DL,0}^E=2.4\times 10^3~\Omega^{-1}$cm$^{-1}$ for the Ti/Ni bilayer and $\lambda_{\alpha-\mathrm{W}}=68\pm16$~nm and $\xi_\mathrm{DL,0}^E=34\times 10^3~\Omega^{-1}$cm$^{-1}$ for the $\alpha$-W/Ni bilayer. The corresponding effective orbital Hall angles in the bulk limit, $\theta_\mathrm{OH,eff}^\mathrm{NM}=\xi_\mathrm{DL,0}^E \rho_\mathrm{NM}^\mathrm{bulk}$, are $\theta_\mathrm{OH,eff}^\mathrm{Ti}=0.13$ and $\theta_\mathrm{OH,eff}^{\alpha-\mathrm{W}}=0.45$, where $\rho_\mathrm{NM}^\mathrm{bulk}$ is the bulk-limit resistivity of the NM layer: $\rho_\mathrm{Ti}^\mathrm{bulk}=54.07$~$\mu\Omega$cm and $\rho_{\alpha-\mathrm{W}}^\mathrm{bulk}=13.30$~$\mu\Omega$cm (see Supplementary Note 11). 
These values are significantly larger than the spin-diffusion length $\lambda_{\mathrm{NM}}^\mathrm{sd}$, the spin Hall conductivity $\sigma_\mathrm{SH}^\mathrm{NM}$, and the spin Hall angle $\theta_\mathrm{SH}^\mathrm{NM}$ of Ti and $\alpha$-W: $\lambda_{\mathrm{Ti}}^\mathrm{sd}\approx 13$~nm, $\sigma_\mathrm{SH}^\mathrm{Ti}= -1.2$ $(\hbar/e)~\Omega^{-1}$cm$^{-1}$, and $\theta_\mathrm{SH}^\mathrm{Ti}=-3.6\times 10^{-4}$ in Ti~\cite{du2014systematic}; $\lambda_{\mathrm{W}}^\mathrm{sd}\approx 1.5$~nm, $\sigma_\mathrm{SH}^{\alpha-\mathrm{W}}=-785$~$(\hbar/e)~\Omega^{-1}$cm$^{-1}$, and $|\theta_\mathrm{SH}^{\alpha-\mathrm{W}}|<0.07$ in $\alpha$-W~\cite{PhysRevMaterials.2.014403,doi:10.1063/1.4753947,PhysRevB.96.241105}. Furthermore, the orbital torque efficiency of $\alpha$-W, $\xi_\mathrm{DL}^E\sim 10^4$ $\Omega^{-1}$cm$^{-1}$, is an order of magnitude larger than that of the Ni/Pt bilayer (see Fig.~\ref{fig2}f and Supplementary Note 12), demonstrating the giant orbital torque.

The observation of the high orbital torque efficiency suggests the existence of a mechanism that generates orbital torques beyond the intrinsic OHE. 
One possible mechanism is an extrinsic OHE originating from the skew scattering, theory of which has not been developed to date. The skew scattering, which relies on disorder scattering, has been shown to lead to exceptionally high anomalous Hall conductivities beyond the intrinsic mechanism~\cite{RevModPhys.82.1539,doi:10.1126/sciadv.abb6003,fujishiro2021giant}. 
This mechanism becomes dominant in the clean limit because the skew-induced Hall conductivity is proportional to the electric conductivity, while the intrinsic Hall conductivity is insensitive to the electric conductivity~\cite{RevModPhys.82.1539}. The high conductivity of the $\alpha$-W layers, $\sim 10^5$~$\Omega^{-1}$cm$^{-1}$, shows that the $\alpha$-W layers are possibly in this clean regime.

To examine the possibility of the extrinsic mechanism, in Fig.~\ref{fig5}e, we plot $\xi_\mathrm{DL}^E$ for all the devices with different $t_\mathrm{NM}$ investigated in this study against the electric conductivity $\sigma_\mathrm{NM}$ of the NM layer in each device. Figure~\ref{fig5}e shows that the high orbital torque efficiency, $\propto T^\mathrm{int}_\mathrm{OH}\sigma_\mathrm{OH}^\mathrm{NM}$, of the W devices increases with $\sigma_\mathrm{NM}$, consistent with the extrinsic mechanism. 
However, the scaling relation between $\xi_\mathrm{DL}^E$ and $\sigma_\mathrm{NM}$ deviates from the linear scaling predicted from the conventional skew scattering theory. This suggests that to understand the result based on the skew scattering, it is necessary to further take into account other possibilities, such as the increase in $T_\mathrm{OH}^\mathrm{int}$ with $t_\mathrm{NM}$ due to the long orbital decay length, suggested by the present experimental results and numerical calculations, and abnormal scaling behavior of the skew scaling mechanism of the OHE. 
The skew scattering scenario is one of the possibilities to explain the experimental observation, and clarifying the exact mechanism remains to be a challenge for future study.

\bigskip
\noindent 
\textbf{Discussion}\\
Over the past three decades, extensive efforts have been directed towards discovering and understanding phenomena arising from spin currents and spin torques, leading to the rapid and exciting development of spintronics. 
\textcolor{black}{In contrast, the exploration of the physics of orbital transport has only just begun, and the fundamental properties of orbital currents and orbital torques have been elusive.}
\textcolor{black}{Thus, in the present work, we focus on revealing the fundamental properties of orbital currents rather than demonstrating and optimizing the ability of orbital currents for specific applications, such as magnetization switching devices. This is the reason why we focus on the characterization of the torque efficiency in the devices based on Ni, which is predicted to show the strongest orbital response among the conventional 3$d$ FMs.}

The present work provides evidence for the orbital response by the systematic measurements of the current-induced torque for the different combinations and thicknesses of the NM and FM layers with the theoretical calculations. In particular, the present work provides unambiguous demonstration of the long-range transport of angular momentum in Ni, which is a unique feature of the orbital transport. 
\textcolor{black}{In metallic FMs, the spin transport length is limited to be less than 1~nm by spin dephasing. The present work demonstrates a means of long-range angular momentum transport in a metallic FM, as well as in NMs, originating in the electronic structure of a material. This result suggests the possibility to realize spintronic functionalities beyond magnetization switching, such as transmitting the signals between different components in the array of diverse elements, which may be termed \textit{orbitronic interconnect}. In a spin-based metallic interconnect, only NMs with weak spin-orbit coupling can be used. In contrast, in the orbital interconnect, even metallic FMs and NMs with strong spin-orbit coupling can be used, providing a new degree of freedom in device design. Further experiments, such as nonlocal transport measurements~\cite{gorbachev2014detecting}, are necessary for accurate determination of the orbital decay length and direct demonstration of the long-range transmission of orbital angular momentum. 
}

The gigantic torque efficiency in the W/Ni devices far exceeds the efficiency of spin-orbitronic devices based not only on Pt but also on exotic materials, such as topological insulators, by an order of magnitude, and the power consumption of the orbital devices can be lower than that of the representative spin-orbitronic devices (see Supplementary Note 12). The present work also demonstrates the efficient torque generation by using Ti, which is light, environment-friendly, abundant on earth, and cheap. 
We also note that the ability of orbital currents to increase the torque efficiency by tuning the FM-layer thickness and materials provides more room for optimizing device parameters, such as the resistance and power consumption. In contrast, spin-based devices do not have this degree of freedom, and the efficiency is fixed by the choice of the spin-Hall layer. This suggests that the tunability of the orbital response offers a unique advantage in device applications.
These results imply that our work has an impact not only on the research in academia but also on the development of devices by industries. 
\textcolor{black}{To realize orbital-based devices, it is important to explore FMs that exhibit strong spin-orbit correlation and are compatible with the current spintronics technology.}

\textcolor{black}{At this stage, orbital torques are not optimized for magnetization switching devices, as thick NM and FM layers are necessary to maximize the orbital torque efficiency. Our results show that the switching power consumption can be further reduced by reducing the orbital transport length. This point has not been recognized previously because the orbital transport has been believed to be short-ranged due to orbital quenching. Although the relaxation mechanism of orbital currents is not clear at this stage, it is possible that nearly degenerate states are responsible for the long-range orbital transport~\cite{go2021long}. This suggests that the orbital decay length in FMs and NMs can be controlled by engineering the band structures, such as by alloying. The FM-layer thickness can also be minimized by using a mechanism of the orbital-torque generation that is different from the mechanism observed in the present study. In this study, the process of the orbital-torque generation is of the second order in the spin-orbit coupling of the FM layer (see Supplementary Note 1). This process is associated with the long-range orbital transport, which is responsible for the torque generation, in the FM layer. We note that orbital torques can also be generated by the injection of orbital currents through a process that is of the first order in the spin-orbit coupling of the FM layer~\cite{PhysRevResearch.2.013177}. In this process, the injected orbital current is converted into a spin current by the spin-orbit coupling in the FM layer. Since the angular momentum responsible for the torque generation is carried by the spin current, the torque efficiency is saturated in the scale of spin dephasing length in this scenario. These two mechanisms are predicted to be sensitive to the band structure of the FM layer, implying that the optimum FM layer thickness can be controlled by material design. We therefore believe that our discovery of the long-range orbital transport and gigantic orbital torque efficiencies provides important information for the material design of orbital-based devices, which will stimulate further experimental and theoretical studies and lead to the fundamental understanding of the physics of orbital currents for practical applications.}

\clearpage

\bigskip\noindent
\textbf{Methods}

\bigskip\noindent\textbf{Devices.} 
The ferromagnet (FM) and nonmagnet (NM) layers in the FM($t_\mathrm{FM}$)/NM($t_\mathrm{NM}$) and NM($t_\mathrm{NM}$)/FM($t_\mathrm{FM}$) structures ($\mathrm{FM}=\mathrm{Ni}$ or Ni$_{81}$Fe$_{19}$, $\mathrm{NM}=\mathrm{Ti}$ or W) were fabricated on SiO$_2$ substrates by radio frequency (RF) magnetron sputtering under 6N-purity-Ar atmosphere. Here, $t_\mathrm{FM}$ and $t_\mathrm{NM}$ represent the thickness of the FM and NM layers, respectively. The surface of the films was covered by 4-nm-thick SiO$_2$. Prior to the film deposition, Ti was sputtered in the chamber (at least 5 min, 0.4~Pa, 120~W) to reduce residual hydrogen and oxygen contents. The resulting base pressure in the chamber was better than $5.0\times 10^{-7}~\text{Pa}$. We used a linear shutter, moving at a constant speed ($<0.05$~mm/s), during the sputtering to vary the thickness of the film in each substrate. 
All sputtering process was performed at room temperature. The materials characterization is described in Supplementary Notes 13-16. The resistivity of the FM layer, measured using the four-probe method, is 17.6~$\rm{\mu\Omega cm}$ for Ni and 41.4~$\rm{\mu\Omega cm}$ for Ni$_{81}$Fe$_{19}$.
Since the self-induced torque is non-negligible only when the FM layer is much thicker than the exchange length~\cite{WangASOT}, we have chosen the thicknesses of the Ni and Ni$_{81}$Fe$_{19}$ layers so that we can neglect the self-induced torque; the thicknesses of the Ni and Ni$_{81}$Fe$_{19}$ layers are thinner than the exchange length in all the devices used in this study, except for the experiment in Fig.~\ref{fig2}d.

For the spin-torque ferromagnetic resonance (ST-FMR) measurement, the films were patterned into rectangular strips with a width of 10~$\mu$m and a length of 150~$\mu$m by conventional photolithography followed by Ar milling. On the edges of the strip, Au(200~nm)/Ti(2~nm) electrodes were deposited by the sputtering and patterned by the photolithography and lift-off technique to form a ground-signal-ground (GSG) contact that guides an RF current into the device.

\bigskip\noindent\textbf{Spin-torque ferromagnetic resonance.} 
For the ST-FMR measurement, an RF current with a frequency of $f$ and a power of $P$ was applied along the longitudinal direction of the device. An in-plane magnetic field $H$ was applied with an angle of $\theta$ from the longitudinal direction of the device. The RF current excites FMR through the current-induced damping-like (DL) and field-like (FL) torques, as well as an Oersted field. Under the FMR, the magnetization precession changes the resistance of the device at the frequency of $f$ due to the anisotropic magnetoresistance (AMR), generating a direct current (DC) voltage through the mixing of the RF current and oscillating resistance~\cite{PhysRevLett.106.036601,fang2011spin}. We measured magnetic field $H$ dependence of the DC voltage $V_\mathrm{DC}$ using a bias tee at room temperature. Here, we determined the RF current flowing in the devices by measuring the resistance change due to Joule heating induced by application of DC and RF currents (see Supplementary Note 3). The determined values of the RF current have been used to obtain the applied electric field.

\bigskip\noindent\textbf{Electric resistivity and crystal structure of W.} 
To characterize the crystal structure of the W layer, we plot thicknesses $t_\mathrm{W}$ dependence of the electric resistivity $\rho_\mathrm{W}$ for W films, as shown in Fig.~\ref{fig3}c. For $t_\mathrm{W}>10$~nm, the $t_\mathrm{W}$ dependence $\rho_\mathrm{W}$ is consistent with $\rho_\mathrm{W}(t_\mathrm{W})=a t_\mathrm{W}^{-1}+\rho_\mathrm{W}^{\text{bulk}}$, where $a t_\mathrm{W}^{-1}$ represents the resistivity due to the surface scattering and $\rho_\mathrm{W}^{\text{bulk}}$ is the resistivity in the bulk limit. The extracted bulk resistivity $\rho_\mathrm{W}^\mathrm{bulk}=9.96$~$\mu\Omega$cm indicates that the W layer is low-resistivity $\alpha$-W in this thickness range. This result is supported by X-ray diffraction measurements (see Supplementary Note 14). By decreasing the W thickness, the measured resistivity $\rho_\mathrm{W}$ deviates from the fitting, which indicates that the W layer with $t_\mathrm{W}<10$~nm is a mixture of low-resistivity $\alpha$-phase and high-resistivity $\beta$-phase~\cite{doi:10.1116/1.4936261}.

\bigskip\noindent
\textbf{Data availability}\\
The data that support the findings of this study are available from the corresponding author upon reasonable request.

\clearpage

\textbf{References}\\

\begin{thebibliography}{10}
\expandafter\ifx\csname url\endcsname\relax
  \def\url#1{\texttt{#1}}\fi
\expandafter\ifx\csname urlprefix\endcsname\relax\def\urlprefix{URL }\fi
\providecommand{\bibinfo}[2]{#2}
\providecommand{\eprint}[2][]{\url{#2}}

\bibitem{wolf2001spintronics}
\bibinfo{author}{Wolf, S.} \emph{et~al.}
\newblock \bibinfo{title}{Spintronics: a spin-based electronics vision for the
  future}.
\newblock \emph{\bibinfo{journal}{Science}} \textbf{\bibinfo{volume}{294}},
  \bibinfo{pages}{1488--1495} (\bibinfo{year}{2001}).

\bibitem{Zutic}
\bibinfo{author}{\ifmmode \check{Z}\else \v{Z}\fi{}uti\ifmmode~\acute{c}\else
  \'{c}\fi{}, I.}, \bibinfo{author}{Fabian, J.} \& \bibinfo{author}{Das~Sarma,
  S.}
\newblock \bibinfo{title}{Spintronics: Fundamentals and applications}.
\newblock \emph{\bibinfo{journal}{Rev. Mod. Phys.}}
  \textbf{\bibinfo{volume}{76}}, \bibinfo{pages}{323--410}
  (\bibinfo{year}{2004}).

\bibitem{RevModPhys.80.1517}
\bibinfo{author}{Fert, A.}
\newblock \bibinfo{title}{Nobel lecture: Origin, development, and future of
  spintronics}.
\newblock \emph{\bibinfo{journal}{Rev. Mod. Phys.}}
  \textbf{\bibinfo{volume}{80}}, \bibinfo{pages}{1517--1530}
  (\bibinfo{year}{2008}).

\bibitem{hoffmann2013spin}
\bibinfo{author}{Hoffmann, A.}
\newblock \bibinfo{title}{{Spin Hall effects in metals}}.
\newblock \emph{\bibinfo{journal}{IEEE Trans. Magn.}}
  \textbf{\bibinfo{volume}{49}}, \bibinfo{pages}{5172--5193}
  (\bibinfo{year}{2013}).

\bibitem{RevModPhys.87.1213}
\bibinfo{author}{Sinova, J.}, \bibinfo{author}{Valenzuela, S.~O.},
  \bibinfo{author}{Wunderlich, J.}, \bibinfo{author}{Back, C.~H.} \&
  \bibinfo{author}{Jungwirth, T.}
\newblock \bibinfo{title}{{Spin Hall effects}}.
\newblock \emph{\bibinfo{journal}{Rev. Mod. Phys.}}
  \textbf{\bibinfo{volume}{87}}, \bibinfo{pages}{1213--1260}
  (\bibinfo{year}{2015}).

\bibitem{RevModPhys.91.035004}
\bibinfo{author}{Manchon, A.} \emph{et~al.}
\newblock \bibinfo{title}{Current-induced spin-orbit torques in ferromagnetic
  and antiferromagnetic systems}.
\newblock \emph{\bibinfo{journal}{Rev. Mod. Phys.}}
  \textbf{\bibinfo{volume}{91}}, \bibinfo{pages}{035004}
  (\bibinfo{year}{2019}).

\bibitem{ryu2020current}
\bibinfo{author}{Ryu, J.}, \bibinfo{author}{Lee, S.}, \bibinfo{author}{Lee,
  K.-J.} \& \bibinfo{author}{Park, B.-G.}
\newblock \bibinfo{title}{Current-induced spin-orbit torques for spintronic
  applications}.
\newblock \emph{\bibinfo{journal}{Adv. Mater.}} \textbf{\bibinfo{volume}{32}},
  \bibinfo{pages}{1907148} (\bibinfo{year}{2020}).

\bibitem{PhysRevLett.95.066601}
\bibinfo{author}{Bernevig, B.~A.}, \bibinfo{author}{Hughes, T.~L.} \&
  \bibinfo{author}{Zhang, S.-C.}
\newblock \bibinfo{title}{Orbitronics: The intrinsic orbital current in
  $p$-doped silicon}.
\newblock \emph{\bibinfo{journal}{Phys. Rev. Lett.}}
  \textbf{\bibinfo{volume}{95}}, \bibinfo{pages}{066601}
  (\bibinfo{year}{2005}).

\bibitem{PhysRevB.77.165117}
\bibinfo{author}{Tanaka, T.} \emph{et~al.}
\newblock \bibinfo{title}{{Intrinsic spin Hall effect and orbital Hall effect
  in $4d$ and $5d$ transition metals}}.
\newblock \emph{\bibinfo{journal}{Phys. Rev. B}} \textbf{\bibinfo{volume}{77}},
  \bibinfo{pages}{165117} (\bibinfo{year}{2008}).

\bibitem{PhysRevLett.102.016601}
\bibinfo{author}{Kontani, H.}, \bibinfo{author}{Tanaka, T.},
  \bibinfo{author}{Hirashima, D.~S.}, \bibinfo{author}{Yamada, K.} \&
  \bibinfo{author}{Inoue, J.}
\newblock \bibinfo{title}{{Giant orbital Hall effect in transition metals:
  Origin of large spin and anomalous Hall effects}}.
\newblock \emph{\bibinfo{journal}{Phys. Rev. Lett.}}
  \textbf{\bibinfo{volume}{102}}, \bibinfo{pages}{016601}
  (\bibinfo{year}{2009}).

\bibitem{PhysRevLett.121.086602}
\bibinfo{author}{Go, D.}, \bibinfo{author}{Jo, D.}, \bibinfo{author}{Kim, C.}
  \& \bibinfo{author}{Lee, H.-W.}
\newblock \bibinfo{title}{{Intrinsic spin and orbital Hall effects from orbital
  texture}}.
\newblock \emph{\bibinfo{journal}{Phys. Rev. Lett.}}
  \textbf{\bibinfo{volume}{121}}, \bibinfo{pages}{086602}
  (\bibinfo{year}{2018}).

\bibitem{PhysRevB.98.214405}
\bibinfo{author}{Jo, D.}, \bibinfo{author}{Go, D.} \& \bibinfo{author}{Lee,
  H.-W.}
\newblock \bibinfo{title}{{Gigantic intrinsic orbital Hall effects in weakly
  spin-orbit coupled metals}}.
\newblock \emph{\bibinfo{journal}{Phys. Rev. B}} \textbf{\bibinfo{volume}{98}},
  \bibinfo{pages}{214405} (\bibinfo{year}{2018}).

\bibitem{PhysRevMaterials.5.074407}
\bibinfo{author}{Salemi, L.}, \bibinfo{author}{Berritta, M.} \&
  \bibinfo{author}{Oppeneer, P.~M.}
\newblock \bibinfo{title}{{Quantitative comparison of electrically induced spin
  and orbital polarizations in heavy-metal/$3d$-metal bilayers}}.
\newblock \emph{\bibinfo{journal}{Phys. Rev. Materials}}
  \textbf{\bibinfo{volume}{5}}, \bibinfo{pages}{074407} (\bibinfo{year}{2021}).

\bibitem{PhysRevB.101.161409}
\bibinfo{author}{Canonico, L.~M.}, \bibinfo{author}{Cysne, T.~P.},
  \bibinfo{author}{Molina-Sanchez, A.}, \bibinfo{author}{Muniz, R.~B.} \&
  \bibinfo{author}{Rappoport, T.~G.}
\newblock \bibinfo{title}{{Orbital Hall insulating phase in transition metal
  dichalcogenide monolayers}}.
\newblock \emph{\bibinfo{journal}{Phys. Rev. B}}
  \textbf{\bibinfo{volume}{101}}, \bibinfo{pages}{161409}
  (\bibinfo{year}{2020}).

\bibitem{PhysRevB.101.075429}
\bibinfo{author}{Canonico, L.~M.}, \bibinfo{author}{Cysne, T.~P.},
  \bibinfo{author}{Rappoport, T.~G.} \& \bibinfo{author}{Muniz, R.~B.}
\newblock \bibinfo{title}{{Two-dimensional orbital Hall insulators}}.
\newblock \emph{\bibinfo{journal}{Phys. Rev. B}}
  \textbf{\bibinfo{volume}{101}}, \bibinfo{pages}{075429}
  (\bibinfo{year}{2020}).

\bibitem{PhysRevB.102.035409}
\bibinfo{author}{Bhowal, S.} \& \bibinfo{author}{Satpathy, S.}
\newblock \bibinfo{title}{{Intrinsic orbital and spin Hall effects in monolayer
  transition metal dichalcogenides}}.
\newblock \emph{\bibinfo{journal}{Phys. Rev. B}}
  \textbf{\bibinfo{volume}{102}}, \bibinfo{pages}{035409}
  (\bibinfo{year}{2020}).

\bibitem{PhysRevLett.126.056601}
\bibinfo{author}{Cysne, T.~P.} \emph{et~al.}
\newblock \bibinfo{title}{{Disentangling orbital and valley Hall effects in
  bilayers of transition metal dichalcogenides}}.
\newblock \emph{\bibinfo{journal}{Phys. Rev. Lett.}}
  \textbf{\bibinfo{volume}{126}}, \bibinfo{pages}{056601}
  (\bibinfo{year}{2021}).

\bibitem{PhysRevB.103.085113}
\bibinfo{author}{Sahu, P.}, \bibinfo{author}{Bhowal, S.} \&
  \bibinfo{author}{Satpathy, S.}
\newblock \bibinfo{title}{{Effect of the inversion symmetry breaking on the
  orbital Hall effect: A model study}}.
\newblock \emph{\bibinfo{journal}{Phys. Rev. B}}
  \textbf{\bibinfo{volume}{103}}, \bibinfo{pages}{085113}
  (\bibinfo{year}{2021}).

\bibitem{PhysRevB.103.195309}
\bibinfo{author}{Bhowal, S.} \& \bibinfo{author}{Vignale, G.}
\newblock \bibinfo{title}{{Orbital Hall effect as an alternative to valley Hall
  effect in gapped graphene}}.
\newblock \emph{\bibinfo{journal}{Phys. Rev. B}}
  \textbf{\bibinfo{volume}{103}}, \bibinfo{pages}{195309}
  (\bibinfo{year}{2021}).

\bibitem{PhysRevB.101.121112}
\bibinfo{author}{Bhowal, S.} \& \bibinfo{author}{Satpathy, S.}
\newblock \bibinfo{title}{{Intrinsic orbital moment and prediction of a large
  orbital Hall effect in two-dimensional transition metal dichalcogenides}}.
\newblock \emph{\bibinfo{journal}{Phys. Rev. B}}
  \textbf{\bibinfo{volume}{101}}, \bibinfo{pages}{121112}
  (\bibinfo{year}{2020}).

\bibitem{PhysRevResearch.2.033401}
\bibinfo{author}{Go, D.} \emph{et~al.}
\newblock \bibinfo{title}{Theory of current-induced angular momentum transfer
  dynamics in spin-orbit coupled systems}.
\newblock \emph{\bibinfo{journal}{Phys. Rev. Research}}
  \textbf{\bibinfo{volume}{2}}, \bibinfo{pages}{033401} (\bibinfo{year}{2020}).

\bibitem{PhysRevResearch.2.013177}
\bibinfo{author}{Go, D.} \& \bibinfo{author}{Lee, H.-W.}
\newblock \bibinfo{title}{Orbital torque: Torque generation by orbital current
  injection}.
\newblock \emph{\bibinfo{journal}{Phys. Rev. Research}}
  \textbf{\bibinfo{volume}{2}}, \bibinfo{pages}{013177} (\bibinfo{year}{2020}).

\bibitem{go2021long}
\bibinfo{author}{Go, D.} \emph{et~al.}
\newblock \bibinfo{title}{Long-range orbital transport in ferromagnets}.
\newblock \emph{\bibinfo{journal}{arXiv:2106.07928}}  (\bibinfo{year}{2021}).

\bibitem{PhysRevB.66.014407}
\bibinfo{author}{Stiles, M.~D.} \& \bibinfo{author}{Zangwill, A.}
\newblock \bibinfo{title}{Anatomy of spin-transfer torque}.
\newblock \emph{\bibinfo{journal}{Phys. Rev. B}} \textbf{\bibinfo{volume}{66}},
  \bibinfo{pages}{014407} (\bibinfo{year}{2002}).

\bibitem{PtCo-orbital}
\bibinfo{author}{Chen, X.} \emph{et~al.}
\newblock \bibinfo{title}{{Giant antidamping orbital torque originating from
  the orbital Rashba-Edelstein effect in ferromagnetic heterostructures}}.
\newblock \emph{\bibinfo{journal}{Nat. Commun.}} \textbf{\bibinfo{volume}{9}},
  \bibinfo{pages}{2569} (\bibinfo{year}{2018}).

\bibitem{PhysRevB.103.L020407}
\bibinfo{author}{Kim, J.} \emph{et~al.}
\newblock \bibinfo{title}{{Nontrivial torque generation by orbital angular
  momentum injection in ferromagnetic-metal/Cu/Al$_{2}$O$_{3}$ trilayers}}.
\newblock \emph{\bibinfo{journal}{Phys. Rev. B}}
  \textbf{\bibinfo{volume}{103}}, \bibinfo{pages}{L020407}
  (\bibinfo{year}{2021}).

\bibitem{PhysRevResearch.2.013127}
\bibinfo{author}{Zheng, Z.~C.} \emph{et~al.}
\newblock \bibinfo{title}{{Magnetization switching driven by current-induced
  torque from weakly spin-orbit coupled Zr}}.
\newblock \emph{\bibinfo{journal}{Phys. Rev. Research}}
  \textbf{\bibinfo{volume}{2}}, \bibinfo{pages}{013127} (\bibinfo{year}{2020}).

\bibitem{tazaki2020current}
\bibinfo{author}{Tazaki, Y.} \emph{et~al.}
\newblock \bibinfo{title}{Current-induced torque originating from orbital
  current}.
\newblock \emph{\bibinfo{journal}{arXiv:2004.09165}}  (\bibinfo{year}{2020}).

\bibitem{PhysRevLett.125.177201}
\bibinfo{author}{Ding, S.} \emph{et~al.}
\newblock \bibinfo{title}{Harnessing orbital-to-spin conversion of interfacial
  orbital currents for efficient spin-orbit torques}.
\newblock \emph{\bibinfo{journal}{Phys. Rev. Lett.}}
  \textbf{\bibinfo{volume}{125}}, \bibinfo{pages}{177201}
  (\bibinfo{year}{2020}).

\bibitem{Cr-orbital}
\bibinfo{author}{Lee, S.} \emph{et~al.}
\newblock \bibinfo{title}{{Efficient conversion of orbital Hall current to spin
  current for spin-orbit torque switching}}.
\newblock \emph{\bibinfo{journal}{Commun. Phys.}} \textbf{\bibinfo{volume}{4}},
  \bibinfo{pages}{234} (\bibinfo{year}{2021}).

\bibitem{Ta-orbital}
\bibinfo{author}{Lee, D.} \emph{et~al.}
\newblock \bibinfo{title}{Orbital torque in magnetic bilayers}.
\newblock \emph{\bibinfo{journal}{Nat. Commun.}} \textbf{\bibinfo{volume}{12}},
  \bibinfo{pages}{6710} (\bibinfo{year}{2021}).

\bibitem{choi2021observation}
\bibinfo{author}{Choi, Y.-G.} \emph{et~al.}
\newblock \bibinfo{title}{{Observation of the orbital Hall effect in a light
  metal Ti}}.
\newblock \emph{\bibinfo{journal}{arXiv:2109.14847}}  (\bibinfo{year}{2021}).

\bibitem{PhysRevLett.128.067201}
\bibinfo{author}{Ding, S.} \emph{et~al.}
\newblock \bibinfo{title}{{Observation of the orbital Rashba-Edelstein
  magnetoresistance}}.
\newblock \emph{\bibinfo{journal}{Phys. Rev. Lett.}}
  \textbf{\bibinfo{volume}{128}}, \bibinfo{pages}{067201}
  (\bibinfo{year}{2022}).

\bibitem{PhysRevResearch.4.033037}
\bibinfo{author}{Sala, G.} \& \bibinfo{author}{Gambardella, P.}
\newblock \bibinfo{title}{{Giant orbital Hall effect and orbital-to-spin
  conversion in $3d$, $5d$, and $4f$ metallic heterostructures}}.
\newblock \emph{\bibinfo{journal}{Phys. Rev. Research}}
  \textbf{\bibinfo{volume}{4}}, \bibinfo{pages}{033037} (\bibinfo{year}{2022}).

\bibitem{du2014systematic}
\bibinfo{author}{Du, C.}, \bibinfo{author}{Wang, H.}, \bibinfo{author}{Yang,
  F.} \& \bibinfo{author}{Hammel, P.~C.}
\newblock \bibinfo{title}{{Systematic variation of spin-orbit coupling with
  $d$-orbital filling: Large inverse spin Hall effect in 3$d$ transition
  metals}}.
\newblock \emph{\bibinfo{journal}{Phys. Rev. B}} \textbf{\bibinfo{volume}{90}},
  \bibinfo{pages}{140407} (\bibinfo{year}{2014}).

\bibitem{PhysRevLett.106.036601}
\bibinfo{author}{Liu, L.}, \bibinfo{author}{Moriyama, T.},
  \bibinfo{author}{Ralph, D.~C.} \& \bibinfo{author}{Buhrman, R.~A.}
\newblock \bibinfo{title}{{Spin-torque ferromagnetic resonance induced by the
  spin Hall effect}}.
\newblock \emph{\bibinfo{journal}{Phys. Rev. Lett.}}
  \textbf{\bibinfo{volume}{106}}, \bibinfo{pages}{036601}
  (\bibinfo{year}{2011}).

\bibitem{fang2011spin}
\bibinfo{author}{Fang, D.} \emph{et~al.}
\newblock \bibinfo{title}{Spin-orbit-driven ferromagnetic resonance}.
\newblock \emph{\bibinfo{journal}{Nat. Nanotechnol.}}
  \textbf{\bibinfo{volume}{6}}, \bibinfo{pages}{413--417}
  (\bibinfo{year}{2011}).

\bibitem{PhysRevB.93.180402}
\bibinfo{author}{Emori, S.} \emph{et~al.}
\newblock \bibinfo{title}{Interfacial spin-orbit torque without bulk spin-orbit
  coupling}.
\newblock \emph{\bibinfo{journal}{Phys. Rev. B}} \textbf{\bibinfo{volume}{93}},
  \bibinfo{pages}{180402(R)} (\bibinfo{year}{2016}).

\bibitem{PhysRevApplied.15.L031001}
\bibinfo{author}{Zhu, L.} \& \bibinfo{author}{Buhrman, R.~A.}
\newblock \bibinfo{title}{{Absence of significant spin-current generation in
  Ti/Fe-Co-B bilayers with strong interfacial spin-orbit coupling}}.
\newblock \emph{\bibinfo{journal}{Phys. Rev. Applied}}
  \textbf{\bibinfo{volume}{15}}, \bibinfo{pages}{L031001}
  (\bibinfo{year}{2021}).

\bibitem{WangASOT}
\bibinfo{author}{Wang, W.} \emph{et~al.}
\newblock \bibinfo{title}{Anomalous spin-orbit torques in magnetic single-layer
  films}.
\newblock \emph{\bibinfo{journal}{Nature Nanotechnol.}}
  \textbf{\bibinfo{volume}{14}}, \bibinfo{pages}{819--824}
  (\bibinfo{year}{2019}).

\bibitem{PhysRevB.96.241105}
\bibinfo{author}{Sui, X.} \emph{et~al.}
\newblock \bibinfo{title}{{Giant enhancement of the intrinsic spin Hall
  conductivity in $\ensuremath{\beta}$-tungsten via substitutional doping}}.
\newblock \emph{\bibinfo{journal}{Phys. Rev. B}} \textbf{\bibinfo{volume}{96}},
  \bibinfo{pages}{241105} (\bibinfo{year}{2017}).

\bibitem{PhysRevMaterials.2.014403}
\bibinfo{author}{Wang, T.-C.}, \bibinfo{author}{Chen, T.-Y.},
  \bibinfo{author}{Wu, C.-T.}, \bibinfo{author}{Yen, H.-W.} \&
  \bibinfo{author}{Pai, C.-F.}
\newblock \bibinfo{title}{{Comparative study on spin-orbit torque efficiencies
  from W/ferromagnetic and W/ferrimagnetic heterostructures}}.
\newblock \emph{\bibinfo{journal}{Phys. Rev. Materials}}
  \textbf{\bibinfo{volume}{2}}, \bibinfo{pages}{014403} (\bibinfo{year}{2018}).

\bibitem{fleur}
\bibinfo{howpublished}{\url{http://www.flapw.de}}.

\bibitem{doi:10.1063/1.4753947}
\bibinfo{author}{Pai, C.-F.} \emph{et~al.}
\newblock \bibinfo{title}{{Spin transfer torque devices utilizing the giant
  spin Hall effect of tungsten}}.
\newblock \emph{\bibinfo{journal}{Appl. Phys. Lett.}}
  \textbf{\bibinfo{volume}{101}}, \bibinfo{pages}{122404}
  (\bibinfo{year}{2012}).

\bibitem{RevModPhys.82.1539}
\bibinfo{author}{Nagaosa, N.}, \bibinfo{author}{Sinova, J.},
  \bibinfo{author}{Onoda, S.}, \bibinfo{author}{MacDonald, A.~H.} \&
  \bibinfo{author}{Ong, N.~P.}
\newblock \bibinfo{title}{{Anomalous Hall effect}}.
\newblock \emph{\bibinfo{journal}{Rev. Mod. Phys.}}
  \textbf{\bibinfo{volume}{82}}, \bibinfo{pages}{1539--1592}
  (\bibinfo{year}{2010}).

\bibitem{doi:10.1126/sciadv.abb6003}
\bibinfo{author}{Yang, S.-Y.} \emph{et~al.}
\newblock \bibinfo{title}{{Giant, unconventional anomalous Hall effect in the
  metallic frustrated magnet candidate, KV$_3$Sb$_5$}}.
\newblock \emph{\bibinfo{journal}{Sci. Adv.}} \textbf{\bibinfo{volume}{6}},
  \bibinfo{pages}{eabb6003} (\bibinfo{year}{2020}).

\bibitem{fujishiro2021giant}
\bibinfo{author}{Fujishiro, Y.} \emph{et~al.}
\newblock \bibinfo{title}{{Giant anomalous Hall effect from spin-chirality
  scattering in a chiral magnet}}.
\newblock \emph{\bibinfo{journal}{Nat. Commun.}} \textbf{\bibinfo{volume}{12}},
  \bibinfo{pages}{317} (\bibinfo{year}{2021}).

\bibitem{gorbachev2014detecting}
\bibinfo{author}{Gorbachev, R.~V.} \emph{et~al.}
\newblock \bibinfo{title}{Detecting topological currents in graphene
  superlattices}.
\newblock \emph{\bibinfo{journal}{Science}} \textbf{\bibinfo{volume}{346}},
  \bibinfo{pages}{448--451} (\bibinfo{year}{2014}).

\bibitem{doi:10.1116/1.4936261}
\bibinfo{author}{Lee, J.-S.}, \bibinfo{author}{Cho, J.} \&
  \bibinfo{author}{You, C.-Y.}
\newblock \bibinfo{title}{Growth and characterization of $\alpha$ and
  $\beta$-phase tungsten films on various substrates}.
\newblock \emph{\bibinfo{journal}{J. Vac. Sci. Technol.}}
  \textbf{\bibinfo{volume}{34}}, \bibinfo{pages}{021502}
  (\bibinfo{year}{2016}).

\bibitem{li2013ni}
\bibinfo{author}{Li, W.} \emph{et~al.}
\newblock \bibinfo{title}{{Ni layer thickness dependence of the interface
  structures for Ti/Ni/Ti trilayer studied by X-ray standing waves}}.
\newblock \emph{\bibinfo{journal}{ACS Appl. Mater. Interfaces.}}
  \textbf{\bibinfo{volume}{5}}, \bibinfo{pages}{404--409}
  (\bibinfo{year}{2013}).

\end{thebibliography}

\clearpage

\bigskip\noindent
Correspondence and requests for materials should be addressed to K.A. (ando@appi.keio.ac.jp)
\\

\bigskip\noindent
\textbf{Acknowledgements}\\
This work was supported by JSPS KAKENHI (Grant Number 22H04964, 19H00864), JST FOREST Program (Grant Number JPMJFR2032), Canon Foundation, and Spintronics Research Network of Japan (Spin-RNJ). H.H. is supported by JSPS Grant-in-Aid for Research Fellowship for Young Scientists (DC1) (Grant Number 20J20663). D.J. and H.-W.L. acknowledge the financial support by the SSTF (Grant No. BA-1501-51). D.G. and Y.M. acknowledge Deutsche Forschungsgemeinschaft (DFG, German Research Foundation) - TRR 173/2 - 268565370 - Spin+X (Project A11), and TRR 288 - 422213477 (Project B06), for funding. We also gratefully acknowledge the Jülich Supercomputing Centre and RWTH Aachen University for providing computational resources under projects jiff40 and jara0062.

\bigskip\noindent
\textbf{Competing interest}\\
The authors declare no competing interests.

\bigskip\noindent
\textbf{Author contributions}\\
H.H., T.G., and S.H. fabricated the devices, performed the experiments, and analyzed the data. D.J. performed the numerical calculations with the help from H.W.L., D.G., and Y.M. K.A. wrote the manuscript with the help from H.H., D.J., H.W.L, D.G, Y.M, T.G., and S.H. All authors discussed results and reviewed the manuscript. K.A. supervised the study.

\clearpage

\end{document}